\documentclass{aastex631}
\usepackage{amsmath}
\usepackage{graphicx}
\usepackage{threeparttable}
\usepackage{txfonts}
\usepackage{soul}
\usepackage{bm} 
\usepackage{tabularx} 
\usepackage{booktabs}
\usepackage{varwidth}
\usepackage{graphicx}

\shorttitle{The Role of Plasma Lensing in FRBs}
\shortauthors{Li et al.}

\begin{document}
\title{The Role of Plasma Lensing in Fast Radio Bursts}

\correspondingauthor{Fa-Yin Wang}
\email{fayinwang@nju.edu.cn}

\author[0009-0007-3326-7827]{Rui-Nan Li}
\affiliation{School of Astronomy and Space Science, Nanjing University Nanjing 210023, China}

\author{Yu-Bin Wang}
\affiliation{School of Physics and Electronic Engineering, Sichuan University of Science \& Engineering, Zigong 643000, China}

\author{Shuang-Xi Yi}
\affiliation{School of Physics and Physical Engineering, Qufu Normal University, Qufu 273165, China}

\author{Xia Zhou}
\affiliation{State Key Laboratory of Radio Astronomy and Technology, Xinjiang Astronomical Observatory, CAS, 150 Science 1-Street, Urumqi, Xinjiang, 830011, China}
\affiliation{Xinjiang Key Laboratory of Radio Astrophysics, 150 Science1-Street, Urumqi, China}

\author[0000-0003-4157-7714]{Fa-Yin Wang}
\affiliation{School of Astronomy and Space Science, Nanjing University Nanjing 210023, China}
\affiliation{Key Laboratory of Modern Astronomy and Astrophysics (Nanjing University) Ministry of Education, China}

\begin{abstract}
Growing evidence indicates that some fast radio bursts (FRBs) reside in dense, magneto-ionic environments where extrinsic propagation effects can substantially reshape the observed signal. Within a 1D Gaussian plasma-lens framework, we show that small, monotonic variations in the incidence angle of the FRB wavefront naturally generate both downward and upward sub-burst frequency drifts. We further demonstrate that distinct lensed paths that probe different rotation measures (RMs), can produce orthogonal polarization-angle (PA) jumps at gigahertz frequencies. In this picture, a $\sim 90^\circ$ PA transition requires only a modest RM contrast of order a few $\times10~\rm{rad~m^{-2}}$ between the multiple images. The chromatic activity of FRB 20180916B—earlier and narrower activity windows at higher frequencies—can be explained as preferential magnification near the outer caustic. Finally, the intrinsic resolution of a plasma lens provides an upper limit on the transverse emission size: lenses located close to the source yield magnetospheric-scale constraints and offer a practical means of discriminating between inner- and outer-magnetospheric emission scenarios. These results suggest that plasma lensing could account for multiple complex observational features of FRBs and may play a non-negligible role in modulating their observable properties.
\end{abstract}


\section{Introduction} \label{sec:intro}
Fast radio bursts (FRBs) are millisecond-duration, high-brightness radio flashes of unknown origin discovered in 2007 \citep{Lorimer2007, Xiao2021, Zhang2023Review}. Most are apparently non-repeating, while $\sim$60 sources repeat \citep{Spitler2016, CHIME/FRB2019, r2}. Their inferred brightness temperatures, up to $\sim10^{35}$ K, require a coherent emission mechanism. Although the central engine remains uncertain, the magnetar scenario is strongly supported by the association of FRB 20200428 with the Galactic magnetar SGR 1935+2154 \citep{CHIME/FRB2020, Bochenek2020}. Emission models are often grouped into close-in (inner-magnetospheric) and far-away (outer-magnetospheric) categories. Critically, the physical size of the emission region is key to distinguishing between these two classes.
Long-term, wide-band monitoring of active repeaters (e.g., FRB 20121102A, FRB 20180916B, FRB 20190520B, FRB 20201124A) now enables year-scale studies of their local environments \citep{Michilli2018, Mckinven2023b, Anna-Thomas2023, Ocker2023, Xu2022, Feng2025}. These observations indicate that the local environments of some repeating FRBs are dynamic, magneto-ionic media characterized by temporal variations in dispersion measure (DM), rotation measure (RM) and scattering times, indicative of multi-path propagation effects. Proposed environments include magnetar wind nebulae \citep{Margalit2018, Yang2019, Zhao2021,Bhattacharya2024,Rahaman2025}, supernova remnants (SNRs) \citep{Piro2016, Yang2017, Piro2018, Zhao2021b, Wang2025}, companion winds and disks \citep{Wang2022, Anna-Thomas2023, Zhao2023,Rajwade2023,Shan2025}, and even regions near supermassive black holes \citep{Zhang2018b, Li2023b}.

Refraction by plasma regions of enhanced or depleted electron density can deflect FRB signals from their initial paths. Such effect has been found in the Crab pulsar, where dispersive delays and strong intensity modulations are attributed to nebular filaments \citep{Backer2000, Smith2011}. In several ``black widow" systems, e.g., PSR B1957+20, PSR J2051$-$0827 \citep{Main2018, Lin2021} and one “redback” system, PSR 1744-24A \citep{Bilous2019}, pronounced, frequency-dependent flux magnifications and DM variations near eclipse phases are naturally explained by plasma lensing in the ionized companion wind. Similarly, plasma lensing can imprint characteristic modulations on the observed properties of FRBs. For FRB 20121102A, complex bifurcated burst structures and deviations from the cold-plasma DM law can be reproduced with $\sim$10-au Gaussian lenses in the host galaxy \citep{Platts2021}. For FRB 20201124A, an abrupt transition to quiescence following a burst-rate surge has been interpreted as defocusing by an over-dense lens \citep{ChenXC2024}. Moreover, the DM of FRB 20180916B shows a bimodal frequency dependence, with the two modes exhibiting opposite trends, a behavior that can be naturally reproduced within the plasma-lensing framework \citep{WangYB2023}.

Beyond these case studies, theory predicts additional lensing signatures. Exponential plasma clumps can produce multiple imaging and ``inverse" frequency–time structures, yielding frequency-dependent delays and apparent DM deviations akin to those observed in repeaters \citep{Er2018, WangYB2022, Er2022}. Magnetized plasma can impart additional rotation of the linear polarization and partially convert linear to circular polarization, beyond standard Faraday rotation and Faraday conversion \citep{Er2023}.

In this work, we demonstrate that a one-dimensional (1D) Gaussian plasma-lens naturally reproduces several observed FRB phenomena.
In Section \ref{sec:Drift}, we demonstrate that within a caustic region a monotonic change in the incidence angle of the FRB wavefront across the lens naturally yields systematic downward or upward drifts, depending on the direction of that change.
In Section \ref{sec:PA}, we show that plausible parameters can reproduce millisecond-timescale orthogonal polarization-angle jumps.
In Section \ref{sec:chromatic} we show that plasma lensing can also generate the observed chromatic behavior under appropriate conditions.
Determining the emission-region size is crucial for identifying the physical origin of FRBs. 
In Section \ref{sec:emission size}, we outline a confirmed plasma-lensing event can be used to constrain and potentially map the emission-region size. We conclude with a discussion in Section \ref{sec:dis} and a summary in Section \ref{sec:con}.

\section{1D Gaussian Plasma lensing} \label{sec:plasma lensing}
\begin{figure} 
\centering
\includegraphics[width=120 mm]{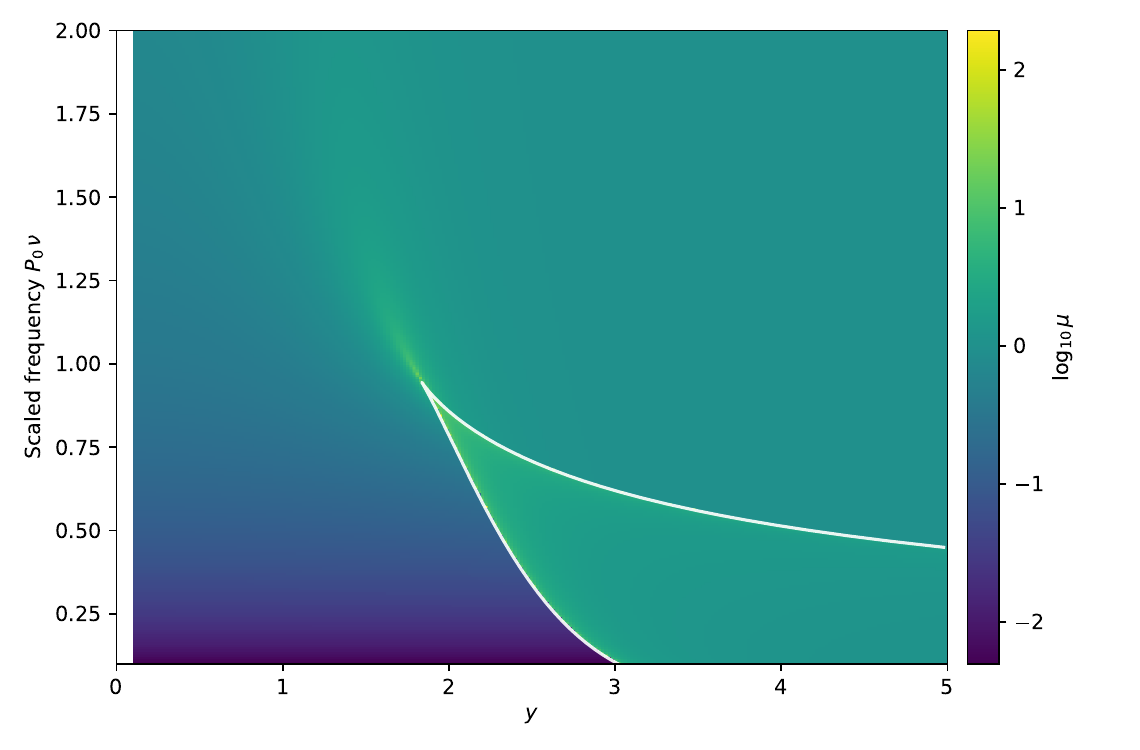}
\caption{The flux magnification map for the 1D Gaussian plasma lens with the color bar scaled logarithmically. The white line represents the caustic curve. $y = 0$ represents the center of the Gaussian distribution of electron column density. So the regions where $y>0$ and $y<0$ are symmetric. For simplicity, only the region where $y>0$ is shown in the figure.
}
\label{Fig1}
\end{figure}

In this work, we adopt the 1D coordinate system centered on the lens and employ the thin-lens approximation, which enables the projection of the lensing mass and plasma distribution onto the coordinate plane.
When light rays emitted from an finitely distant source pass through a region with significant fluctuations in electron number density, the rays are deflected by an angle $\alpha$, which is related to the phase change $\delta \phi_{\mathrm{les}}$ as \citep{Schneider1992, Wagner2020}
\begin{equation}\label{equ:1}
{\alpha}=\frac{1}{k}{\nabla}\delta\phi_{\mathrm{les}},
\end{equation}
where $k$ is the wave number. 
As shown in panel (a) of Figure \ref{Fig2}, for a fixed source position, the source is initially located at an undisturbed angular position $\beta$. After the light ray is deflected, an image of the source appears at position $\theta$. The relationship between $\beta$ and $\theta$ is
\begin{equation}
{\beta}=\theta-\frac{D_{ds}}{D_s}\alpha,
\end{equation}
where $D_d$, $D_s$ are the angular-diameter distances from observer to lens, observer to source, respectively. Alternatively, in terms of the coordinates on the image plane $X= D_d\theta$ and the source plane $Y=D_s \beta$:
\begin{equation}\label{equ:2}
{Y}=\frac{D_s}{D_d}{X}-D_{ds}\alpha(X),
\end{equation}
where $D_{ds}$ is the angular-diameter distances from lens to source. The phase change $\delta \phi_{\mathrm{les}}$ is a function of $X$, expressed as:
\begin{equation}\label{equ:3}
\delta \phi_{\mathrm{les}}(X)=-\lambda r_e N_{\mathrm{e}}(X),
\end{equation}
where $\lambda$, $r_e$, $N_{\mathrm{e}}(X)$ denote the wave length, classical electron radius and electron surface overdensity in the lens plane. Substituting equation (\ref{equ:3}) into equation (\ref{equ:1}), and insert the result into equation (\ref{equ:2}), we can rewrite the lens equation as
\begin{equation}\label{equ:4}
{Y}=\frac{D_s}{D_d}{X}-\frac{c^2r_e}{2 \pi \nu^2}D_{ds}\nabla_X N_{\mathrm{e}}(X),
\end{equation}
where $\nu$ is the frequency of light.
Applying a dimensionless scaling with the characteristic lens width $a$, we rewrite the lens equation in terms of the normalized coordinates $y=\frac{D_d}{D_s}\frac{Y}{a}$, $x=\frac{X}{a}$ as
\begin{equation}\label{equ:5}
{y}={x}-\alpha = {x}- \nabla_x \psi({x}),
\end{equation}
where $y$ and $x$ are the normalized angular coordinates in the source plane and image plane respectively, and $\nabla_x$ denotes the gradient operator on the image plane (see \cite{Wagner2020} for details). The effective lensing potential $\psi$ takes the form
\begin{equation}\label{equ:6}
\psi(x) =  \frac{D_{d s} D_d}{D_s} \frac{c^2 r_e N_{\mathrm{e}}(x)}{2 \pi a^2 \nu^2},
\end{equation}
For a geometrically thin plasma lens, the column density can be approximated as $N_{\mathrm{e}}(x) \approx \mathrm{DM_e}(x)$, where $\rm{DM}_{\rm{e}}$ denotes the dispersion measure contributed locally by the lens along the normalized coordinate $x$. 
\cite{ChenXC2024} introduced a frequency-independent parameter, which is solely determined by the physical properties of the lens, to represent the effective potential
\begin{equation}\label{equ:7}
\frac{1}{P_0^2} = \frac{D_{d s} D_d}{D_s} \frac{c^2 r_e N_0}{2 \pi a^2},
\end{equation}
where $N_0$ denotes the maximum electron column density. We can rewrite equation (\ref{equ:5}) as
\begin{equation}\label{equ:8}
{y}={x}-\alpha = {x}- \frac{1}{P_0^2 \nu^2 }\nabla_x n_e({x}),
\end{equation}
where $n_{\mathrm{e}}(x)$ is the dimensionless density profile. 
The 1D Gaussian lens is a widely used approximation for studying plasma lensing effects in various astrophysical contexts \citep{Clegg1992, Backer2000, Smith2011, cordes2017, Main2018, WangYB2022, ChenXC2024}. While two-dimensional (2D) plasma lensing has been investigated in detail \citep{Tuntsov2016, cordes2017}, the 1D approximation remains a robust framework for capturing the essential physics. As the line of sight traverses a 2D structure, the resulting temporal variations are effectively governed by a 1D cross-section of the potential. Consequently, 1D models reproduce the key qualitative features of their 2D counterparts, including the canonical behavior of ``fold" caustics \citep{Berry1980, cordes2017}. Furthermore, astrophysical plasmas (e.g., in the ISM or circum-source environments) are typically highly magnetized, which induces significant anisotropy. Such structures tend to elongate along magnetic field lines into filaments or sheets with extreme axial ratios \citep{Tuntsov2016}, effectively reducing the lensing geometry to a 1D formalism. Therefore, although the 1D Gaussian profile is an idealization of complex, turbulent environments, it is physically justified and sufficient for the scope of this work. In this work, we adopt an overdense plasma lens with a Gaussian electron-density profile, $n_{\mathrm{e}}(x) = \mathrm{exp}(-x_0^2)$. Equation (\ref{equ:8}) thus can be rewritten as
\begin{equation}\label{equ:9}
{y} = {x} + \frac{1}{P_0^2 \nu^2 } 2xe^{-x^2}.
\end{equation}
The flux density of sources is modified by the lensing effect due to the deflection of light rays. The flux magnification $\mu$ is the ratio of observed flux to source flux, and is quantified by the inverse of the determinant of the Jacobian matrix of the lens mapping \citep{ChenXC2024}: 
\begin{equation}\label{equ:10}
\mu=\left|\delta_{i j}-\frac{1}{P_0^2 \nu^2 } \frac{\partial^2 n_e({x})}{\partial x_i \partial x_j}\right|^{-1},
\end{equation}
where $x_i,x_j$ denote two dimensional coordinates in the lens plane. From this equation, it can be inferred that, at certain fixed positions, the flux magnification theoretically approaches infinity at specific frequencies. These positions of divergence in flux magnification correspond to the regions of extreme flux magnification at the caustic plane. In the frequency domain, this is expected to manifest as a two-peak structure in the spectrum, arising at the caustic plane. Under the 1D Gaussian assumption, the caustic in the lens plane satisfies \citep{cordes2017}
\begin{equation}\label{equ:11}
y_c=\frac{2 x^3}{2 x^2-1}.
\end{equation}
The flux magnification for the 1D Gaussian plasma lens is illustrated in Figure \ref{Fig1}. Theoretically, the minimum value of $y_{c,\rm{min}}$ occurs at $y_{c,\rm{min}}= (3/2)^{\frac{3}{2}}$. For any value of $y$ greater than $y_{c,\rm{min}}$, there exists a corresponding caustic region in the frequency domain, with the two boundaries of this region corresponding to the extreme flux magnification. Within this caustic region, multiple images are formed. However, as shown in Figure \ref{Fig1}, the frequency range of the caustic region depends on the properties of the plasma lens $P_0$.  

\section{Frequency drifting patterns} \label{sec:Drift}
The frequency-dependent arrival-time delay is a key observational feature in the morphological studies of FRBs. This behavior has been observed in many active repeaters, typically manifesting in two distinct forms: intra-burst drifting and inter-sub-burst drifting \citep{Zhou2022, ZhangLX2025}. These patterns are likely to arise from the same underlying physical mechanism. Within the magnetospheric framework, the radius-to-frequency mapping has been invoked to explain such behavior \citep{WangWY2019, Lyutikov2020b}. Alternatively, applying standard de-dispersion techniques to bursts that experience frequency-dependent DMs can naturally result in the formation of the “sad/happy trombone” feature \citep{Tuntsov2021, WangYB2025}. Several other models have also been proposed to account for this feature, including those involving shock-driven emission, scintillation, and wave propagation effects \citep{Metzger2019, Gu2020, Simard2020, Rajabi2020}.

When the separation between sub-bursts is comparable to or shorter than the burst duration, they appear blended into a single event, whereas larger separations keep them distinct. If the central frequency of the emission evolves with time, either the apparent single burst or the individual sub-bursts will exhibit a downward or upward frequency drift. In this section, we show that both downward and upward drifting can arise from the plasma lensing effect.

As shown in Figure \ref{Fig1}, intrinsically weak bursts can be strongly magnified and become detectable within the caustic region. In this work, we focus on scenarios in which bursts are observable only when such lensing-induced magnification boosts their flux above the detection threshold. Because the frequency range over which this magnification occurs depends on the transverse incident coordinate $y$, even small temporal variations in the incidence angle $\theta$ naturally translate into an evolution of $y$ with time. 

\begin{figure} 
\centering
\includegraphics[width=\textwidth]{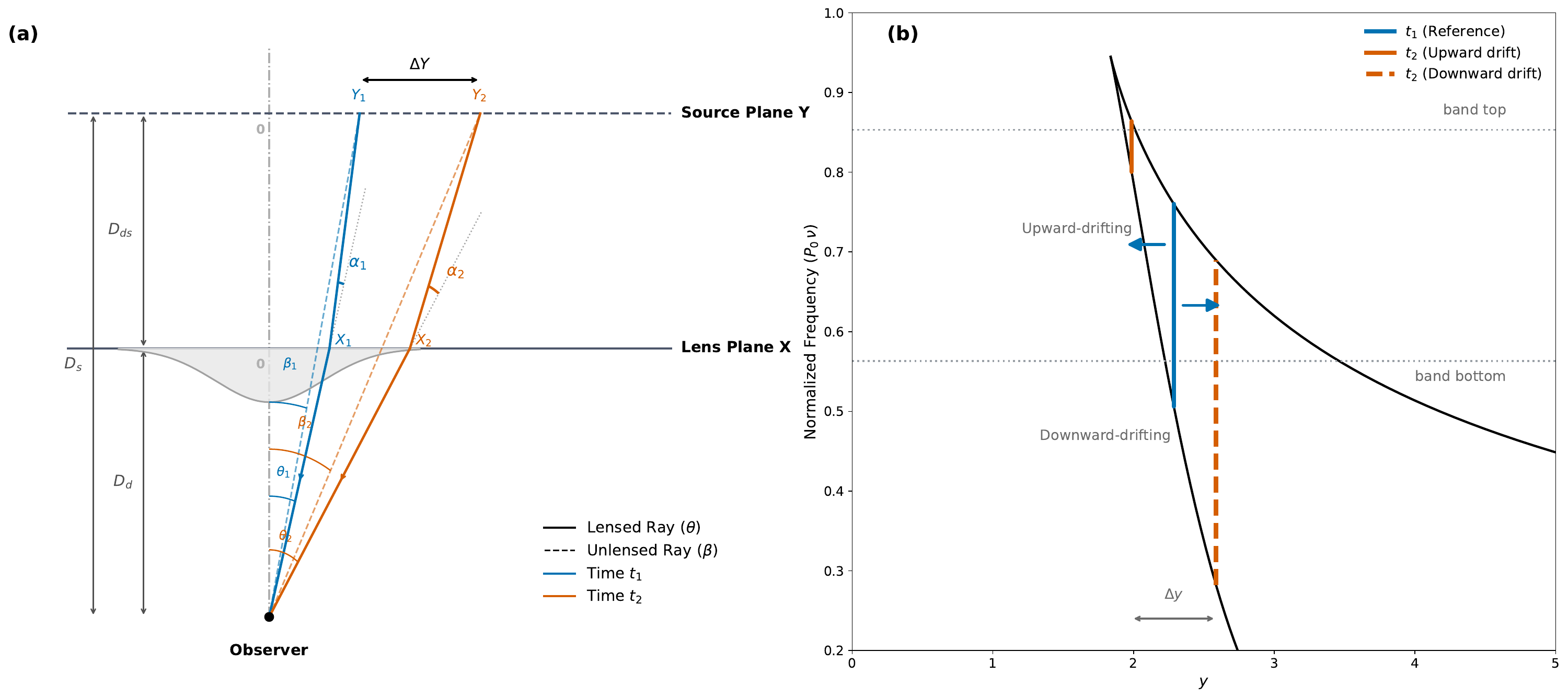}
\caption{Panel (a): Schematic diagram illustrating the geometric setup of the plasma lensing model. The coordinate system is centered on the peak of the lens's plasma density distribution. $Y$ and $X$ denote the physical coordinates on the source plane and lens plane, respectively. $\beta$ and $\theta$ represent the angular positions of the source (unlensed ray) and the image (lensed ray). $\alpha$ are the deflection angles of the incident rays. The dashed lines indicate the unlensed lines of sight, while the solid lines trace the lensed ray paths at two epochs, $t_1$ and $t_2$. This time interval corresponds to a small change in the beam incidence angle $\delta \theta$, resulting in an effective source displacement of $\Delta Y$.
Panel (b): A schematic diagram illustrating how two incident rays, with different values of $y$, are magnified by the lens across different frequency ranges due to the plasma lensing effect. The blue vertical line represents the first observed emission, while the orange dashed and solid vertical lines represent the second observed emission, corresponding to either a positive or negative value of $\Delta y$. The gray dotted lines represent the observing frequency band pass. 
}
\label{Fig2}
\end{figure}

The triggering mechanisms of FRBs are generally believed to involve sudden, high-energy events such as starquakes \citep{WangWY2018, Suvorov2019, Wadiasingh2019, Li2022, Lander2023, WangFY2023, Qu2024, WuQ2025}. Proposed magnetospheric emission mechanisms include coherent curvature radiation and coherent inverse Compton scattering, in which charged bunches streaming along magnetic field lines either directly produce the observed radio emission \citep{Katz2014, Kumar2017, Katz2018, Lu2018, Yang2018, Cooper2021} or upscatter low-frequency waves to FRB frequencies \citep{Zhang2022_ICS, Qu2024}. In these frameworks, different bursts or sub-bursts are naturally associated with charge bunches located at different positions along curved field lines or on neighboring field lines. Because these bunches radiate at different times and at slightly different locations along the curved field lines, the local magnetic-field tangent varies between bursts and consequently the direction of the radiation beam also varies slightly. In the scenario of a rotating beam, the sweeping of the beam across the lens plane from $X_1$ to $X_2$ corresponds to a shift in the effective source coordinate from $Y_1$ to $Y_2$. When this emission propagates toward a near-source plasma screen, subtle variations in the beam direction ($\delta \theta$) translate into differences in the effective normalized source coordinate $\Delta y$. A schematic representation of this scenario is shown in panel (a) of Figure \ref{Fig2}. According to equation (\ref{equ:4}), the relationship between $\Delta y$ and $\delta \theta$ can be expressed as
\begin{equation}\label{equ:12}
\Delta y= \frac{D_d}{D_sa} \delta Y = \frac{D_d}{D_sa} (\delta X + \sigma) = \frac{D_dD_{ds}}{D_sa} \delta \theta + \Sigma,
\end{equation}
where $\sigma$ and $\Sigma$ represent the differential terms arising from the variation in electron column density $N_{\rm{e}}(X)$. Since FRBs are generally extragalactic, the ratio $D_s/D_d \approx 1$. The physical scale of the lens (sub-AU to AU) is significantly smaller than the source-lens distance $D_{ds}$ (typically tens of AU to pc). Consequently, even an extremely small change in the incidence angle, $\delta\theta \sim 10^{-10}$–$10^{-6}$ rad, can result in a significant shift in the dimensionless coordinate, $\Delta y \sim 0.1$. The millisecond-level separations between sub-bursts are sufficient to produce the required $\delta\theta$, thereby giving rise to pronounced frequency-drifting patterns. The sign of $\Delta y$ depends on the sign of $\delta\theta$, which is determined jointly by the rotation of the magnetar and by the specific location at which the emission intersects the lens plane. As shown in the panel (b) of Figure \ref{Fig2}, a negative $\Delta y$ corresponds to an upward-drifting pattern while a positive $\Delta y$ corresponds to a downward-drifting. In this model, bursts at higher frequency bands are expected to exhibit a narrower magnified frequency range. However, this geometric effect is significantly influenced by the telescope's observing frequency bandpass. Additionally, the intrinsic frequency width of each FRB burst can vary randomly. 

\begin{figure} 
\centering
\includegraphics[width=150 mm]{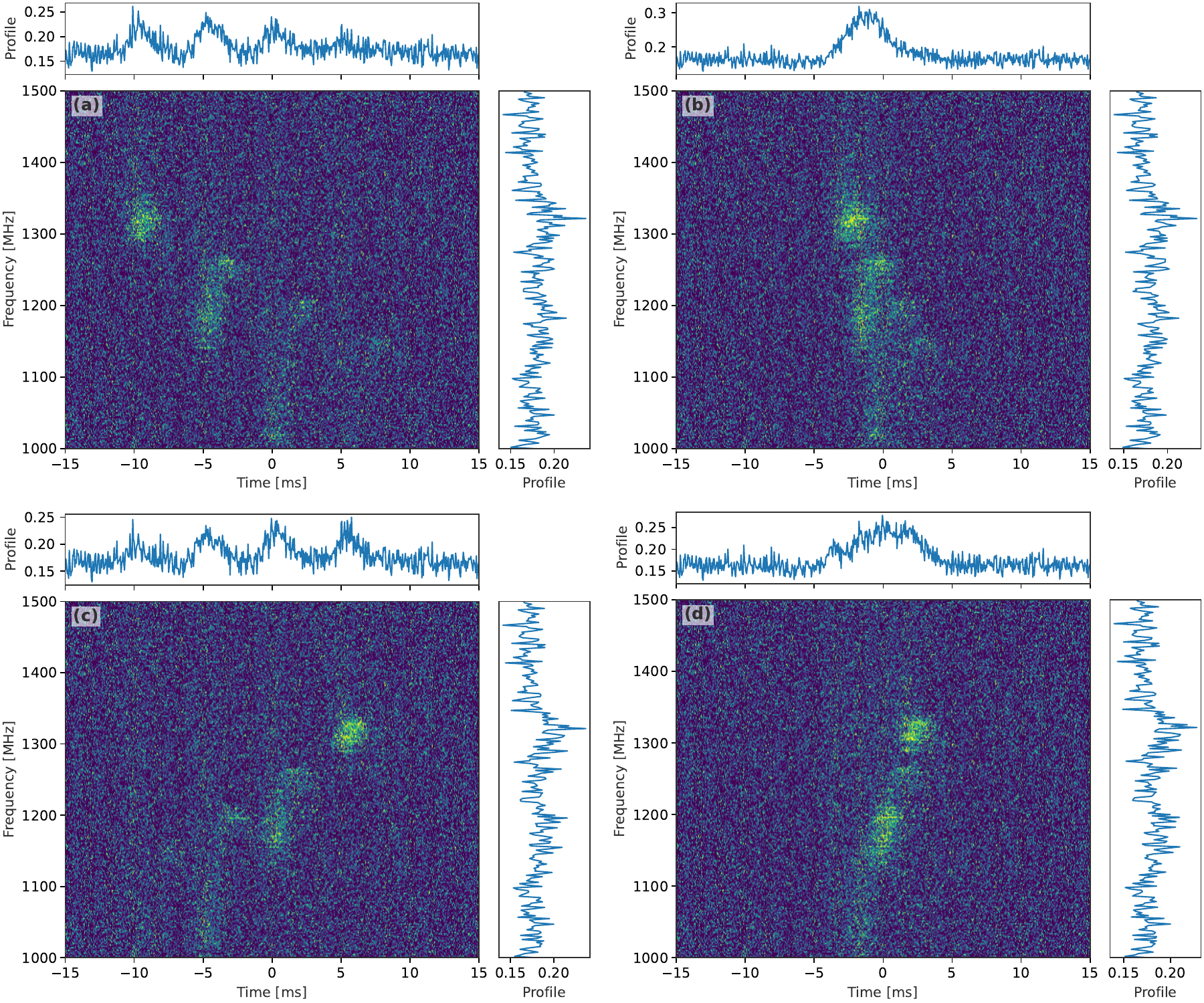}
\caption{Simulated dynamic spectra of four sub-bursts modeled with multi-spot emission and 1D Gaussian plasma lensing. Each panel shows the waterfall (center) with the time-averaged profile (top) and the band-averaged spectrum (right). Frequencies span 1000–1500 MHz with 256 channels. Time spans $-15$ to $+15$ ms with 0.05 ms sampling. Intrinsic radiation is modeled as a Gaussian profile with FWHM of 0.35 GHz, centered at $\nu_0=1.25$ GHz. Lens parameters are defined with a transverse lens scale $a = 7.2 \times 10^{10}$cm, $\rm{DM_l}=1~pc~cm^{-3}$, and $D_{ds} = 6$ AU. Noise is added to the signal after summing the contributions of the individual FRBs. The signal-to-noise ratio is set to 10, and Gaussian noise is added to the simulated signal to match this signal-to-noise ratio. Panel (a) and (b) use $y_0=\{1.9, 2.0, 2.1, 2.2\}$; panel (c) and (d) use $y_0=\{2.2, 2.1, 2.0, 1.9\}$. The time separations between the four sub-bursts are $\Delta t=5\rm{ms}$ for panel (a) and (c), and $\Delta t=1\rm{ms}$ for panel (b) and panel (d).}
\label{Fig3}
\end{figure}

We model dynamic spectra of four sub-bursts in Figure \ref{Fig3} to illustrate how a near-source plasma lens reshapes emission with different inter-burst spacings. Each sub-burst is assigned a distinct lens plane coordinate $y$ and is placed at a prescribed time center. In panels (a) and (c) the spacing is $\Delta t = 5 ~\rm{ms}$, while in panels (b) and (d) it is $\Delta t = 1 ~\rm{ms}$, representing the “resolved sub-bursts” and “apparently single burst” regimes, respectively. The intrinsic spectrum of each sub-burst is a Gaussian envelope centered at $\nu_{ref} = 1250$ MHz, with full-width at half-maximum (FWHM) $f_{FWHM}=0.35~ \rm{GHz}$. To account for beaming, each sub-burst is treated as a bunch of emitters whose radiation angles follow a Gaussian distribution within a $1/f$ angular range (beaming factor $f=100$). The beaming effect smooths and can even remove the sharp frequency spikes that would appear at caustic boundaries. We then apply the lensing transfer function, which incorporates both frequency-dependent flux magnification and geometric delays, to generate the observed spectra. Panels (a) and (b) illustrate downward-drifting structures corresponding to well-separated sub-bursts and an apparently single burst, respectively. While panels (c) and (d) show the same two timing regimes for the upward-drifting case.

\section{Orthogonal polarization angle jump} \label{sec:PA}
Observations of the PA evolution of FRBs reveal two distinct behaviors: either no evolution \citep{Gajjar2018, Michilli2018, CHIME2019, Chawla2020, Nimmo2021+20200120E, Kumar2021}, or varying behavior \citep{Luo2020, Feng2022, Zhang2023, Niu2024, Jiang2024, Bethapudi2025}. For some FRBs, the rotating vector model (RVM) \citep{Radhakrishnan1990} has been applied to study PA evolution, with results supporting a magnetospheric origin for FRBs \citep{Mckinven2025, Liu2025}. Notably, FRB 20201124A exhibits a triggering orthogonal PA jump, a phenomenon commonly seen in pulsars \citep{Manchester1975, Stinebring1984} but not previously observed in other FRBs \citep{Niu2024}. This jump was detected in only three bursts, with a transition time of several milliseconds, out of over 2000 detected bursts. Both intrinsic mechanisms and geometric effects can give rise to this phenomenon, with plasma lensing being one of the possible explanatory pathways \citep{Qu2025}. 

According to the 1D Gaussian plasma-lens mapping $y=(x;\nu)$ defined in Equation (\ref{equ:5}), the Jacobian of the image–source transformation is
\begin{equation}\label{equ:13}
J(x)=\frac{\partial y}{\partial x}=1+ \frac{2}{P_0 \nu}\left(1-2 x^2\right) e^{-x^2}.
\end{equation}

The critical curve is the locus where $J(x)=0$, and its projection into the source plane defines the caustic. In the geometric–optics limit, the flux magnification is $\mu_i = \left| J_i \right|^{-1}$, hence $J\to0$ implies a formal divergence of $\mu$. Multiple imaging arises precisely when the mapping $x \to y$ becomes non-monotonic, i.e. when $J$ changes sign so that a single $y$ admits several solutions $x_i$. Throughout, the subscript $i$ labels distinct images. 

For FRBs embedded in a magneto-ionic environment (e.g. FRB 20201124A), different geometric images produced by a plasma lens traverse distinct portions of the magneto-ionic medium. Consequently, they accumulate different RMs. A minimal linearization in image coordinate $x_i$ is
\begin{equation}\label{equ:14}
\mathrm{RM}_i=\mathrm{RM}_0+\frac{d \mathrm{RM}}{d x} x_i,
\end{equation}
where $d\mathrm{RM}/dx$ parametrizes the local RM gradient across the lens plane. The PA of image $i$ at frequency $v$ then follows the $\lambda^2$ law,
\begin{equation}\label{equ:15}
\chi_i(\nu)=\chi_0+\mathrm{RM}_i \lambda^2, \quad \lambda^2=(c / \nu)^2,
\end{equation}
where $\chi_0$ is the intrinsic PA prior to propagation. $I_i(t,\nu)$ is the specific-intensity contribution of image $i$ including the lensing magnification and geometric delay. $p_i$ denotes the intrinsic polarization fraction. The Stokes parameters for each image are:
\begin{equation}\label{equ:16}
Q_i=p_i I_i \cos \left(2 \chi_i\right), \quad U_i=p_i I_i \sin \left(2 \chi_i\right)
\end{equation}
so that the observed broadband Stokes parameters (non-coherent summation over frequency channels) are
\begin{equation}
\begin{gathered}
Q_{\text {tot }}(t)=\sum_i\left\langle Q_i(t, \nu)\right\rangle_\nu, \quad U_{\text {tot }}(t)=\sum_i\left\langle U_i(t, \nu)\right\rangle_\nu, \\
I_{\text {tot }}(t)=\sum_i\left\langle I_i(t, \nu)\right\rangle_\nu,
\end{gathered}
\end{equation}
where $<...>_v$ denotes averaging over the instrumental channel. The net linear polarization and PA are
\begin{equation}
L(t)=\sqrt{Q_{\mathrm{tot}}^2+U_{\mathrm{tot}}^2}, \quad \chi_{\mathrm{obs}}(t)=\frac{1}{2} \tan^{-1}\left(\frac{U_{\mathrm{tot}}}{Q_{\mathrm{tot}}}\right)
\end{equation}
Because $\chi_i(\nu)$ depends on $\mathrm{RM}_i$, and thus on $x_i$, different images generally have distinct PAs at a given $\nu$. When two images dominate the flux at some time $t$ and their PA difference $\Delta \chi = \chi_1 - \chi_2$ passes near $90^\circ$ producing an apparently sharp PA transition in $\chi_{obs}(t)$. In time-resolved data this typically occurs when the delayed image envelopes overlap within the burst, so that the relative weights $<I_1>$ and $<I_2>$ exchange dominance. Near $\Delta \chi \simeq 90 ^\circ$, the two linearly polarized vectors are nearly orthogonal, for comparable amplitudes the $Q-U$ sum partially cancels, yielding a depression in $L/I$.  

\begin{figure} 
\centering
\includegraphics[width=100 mm]{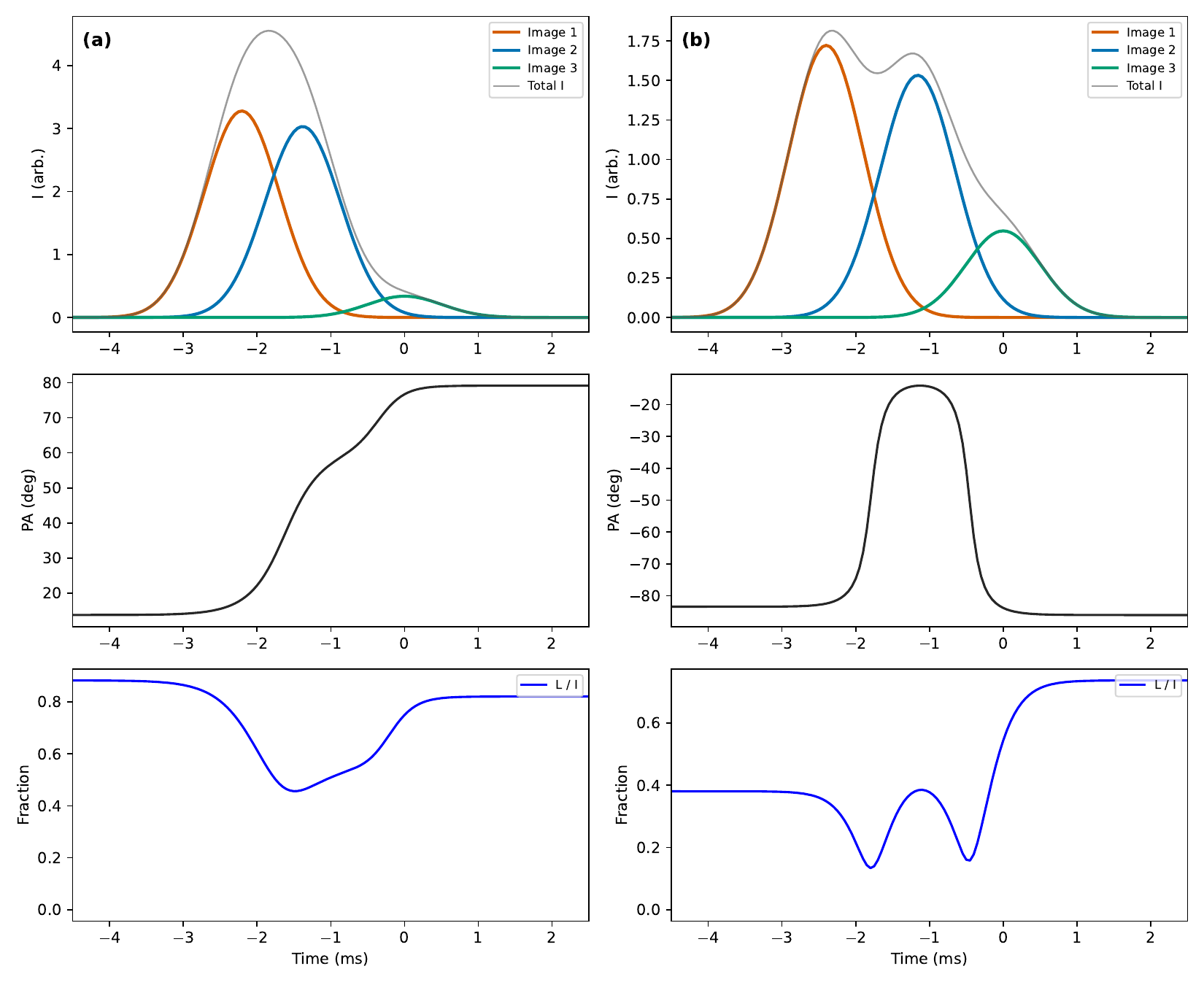}
\caption{Time-domain polarization evolution of a simulated FRB burst lensed by a near-source 1D Gaussian plasma lens for two different lens configurations. The parameters are set to $\alpha_0=2.3, y=2.15$, $\alpha_0=2.0, y=2.20$ and $\rm{RM_0}=50~rad~m^{-2}$ for panel (a) and panel (b), respectively. In each column, the top panel shows the band-averaged intensity of the individual lensed images and the total intensity, the middle panel shows the band-averaged PA, and the bottom panel shows the fractional linear $L/I$. The observing band is 1.15–1.35 GHz with 0.5 MHz channels, and a transverse RM gradient is chosen such that the two brightest images differ by $\Delta \rm{RM}\simeq20~rad~m^{-2}$. 
}
\label{Fig4}
\end{figure}

Figure \ref{Fig4} summarizes the time-domain polarization phenomenology expected from multi-imaging by a near-source 1D Gaussian plasma lens. Top Panel shows how the relative dominance of individual images changes with time. The PA jump in the middle panel coincides with the moment when two images of comparable band-averaged intensity overlap. The different $\mathrm{RM}_i$ makes $\chi_1 - \chi_2 \simeq 90 ^\circ$, so the net Stokes vector ($Q_{\rm{tot}}$, $U_{\rm{tot}}$) rotates rapidly, yielding an apparently sharp PA transition. The strength of each image regulates both the sharpness and the amplitude of the resulting PA variation. In panel (a), only a single prominent PA jump is produced because the third image is too weak to constitute an independent polarization mode. In contrast, panel (b) shows a configuration that yields two substantial PA jumps, closely resembling the multiple orthogonal-mode transitions observed in FRB 20201124A and in many pulsars \citep{Manchester1975, Cordes1978, Backer1980, Radhakrishnan1990, Singh2024, Niu2024}. Bottom panel demonstrates the companion signature: because the two linear-polarization vectors are nearly orthogonal at that instant, their vector sum partially cancels, producing a pronounced depression in $L/I$. 


From Equation (\ref{equ:10}), the difference of PA between two images at frequency $\nu$ is
\begin{equation}
\Delta \chi(\nu)=\left[\mathrm{RM}_1-\mathrm{RM}_2\right] \lambda^2+\left(\chi_{0,1}-\chi_{0,2}\right).
\end{equation}
The required $\Delta \rm{RM}$ for $\Delta \chi$ is
\begin{equation}
\Delta \mathrm{RM} \approx \frac{\Delta \chi}{\lambda^2}.
\end{equation}
At 1.1 GHz, producing a $90^ \circ$ PA jump requires a RM difference of $\Delta \rm{RM} \simeq20~\rm{rad~m^{-2}}$. Variations of this magnitude are commonly seen between adjacent bursts of FRB 20201124A, implying the presence of strong RM gradients in its near-source magneto-ionic environment \citep{Xu2022,Jiang2022}. 
It has been proposed that FRB 20201124A likely resides in a binary system, with the circumstellar material of its companion playing a key role in the complex evolution of RM \citep{Wang2022, LiR2025, XuJW2025, ZhangWL2025}. The clumps in the stellar wind or circumstellar disk, with typical sizes on the order of an AU, can act as plasma lenses \citep{Moffat1994, Puls2006, Chernyakova2006}. The filling factor of these clumps can be as high as 0.1 \citep{Moffat1994, Puls2006,Zhao2023}. As a result, the orthogonal polarization angle jumps observed in FRB 20201124A may be due to plasma lensing effects caused by these clumps in the circumstellar material. 
If a sliding, narrow-band RM fits on either side of the PA jump is performed, a measurable change in the effective RM across the jump is expected and naturally tracks the changing relative weights of the images.

\section{The chromatic active window} \label{sec:chromatic}

FRB 20180916B exhibits a periodicity of 16.33~days, with an active phase lasting approximately $60\%$ of the cycle \citep{CHIME2020180916, Marazuela2021, Bethapudi2023}. Several models have been proposed to explain the physical origin of this periodicity, including the slowly rotating neutron star model \citep{Beniamini2020, LiDZ2021,Lan2024}, free or forced precession \citep{Levin2020, YangH2020, Zanazzi2020, Katz2022_180916}, and binary interaction \citep{Ioka2020, Lyutikov2020, LiQC2021, Wada2021}. Some of these scenarios have been disfavored by recent observations, and the slow-rotation model has emerged as a more plausible explanation \citep{Lan2024, Bethapudi2025}. Notably, the activity window of FRB~20180916B is strongly frequency-dependent: bursts occur earlier and within narrower phase ranges at higher frequencies \citep{Marazuela2021}. This frequency-dependent phase selectivity has been observed across a broad range from 110 MHz to 5 GHz \citep{Pleunis2021_180916, Bethapudi2023}. A geometric model within the dipolar magnetosphere has been proposed to account for this behavior \citep{LiDZ2021}. However, statistical analyses suggest that plasma lensing may also play an important role in shaping the observed properties of FRB 20180916B \citep{WangYB2023}. In this section, we demonstrate that the plasma lensing effect can naturally reproduce the chromatic nature of the active window within the slow-rotation scenario.

\begin{figure} 
\centering
\includegraphics[width=150 mm]{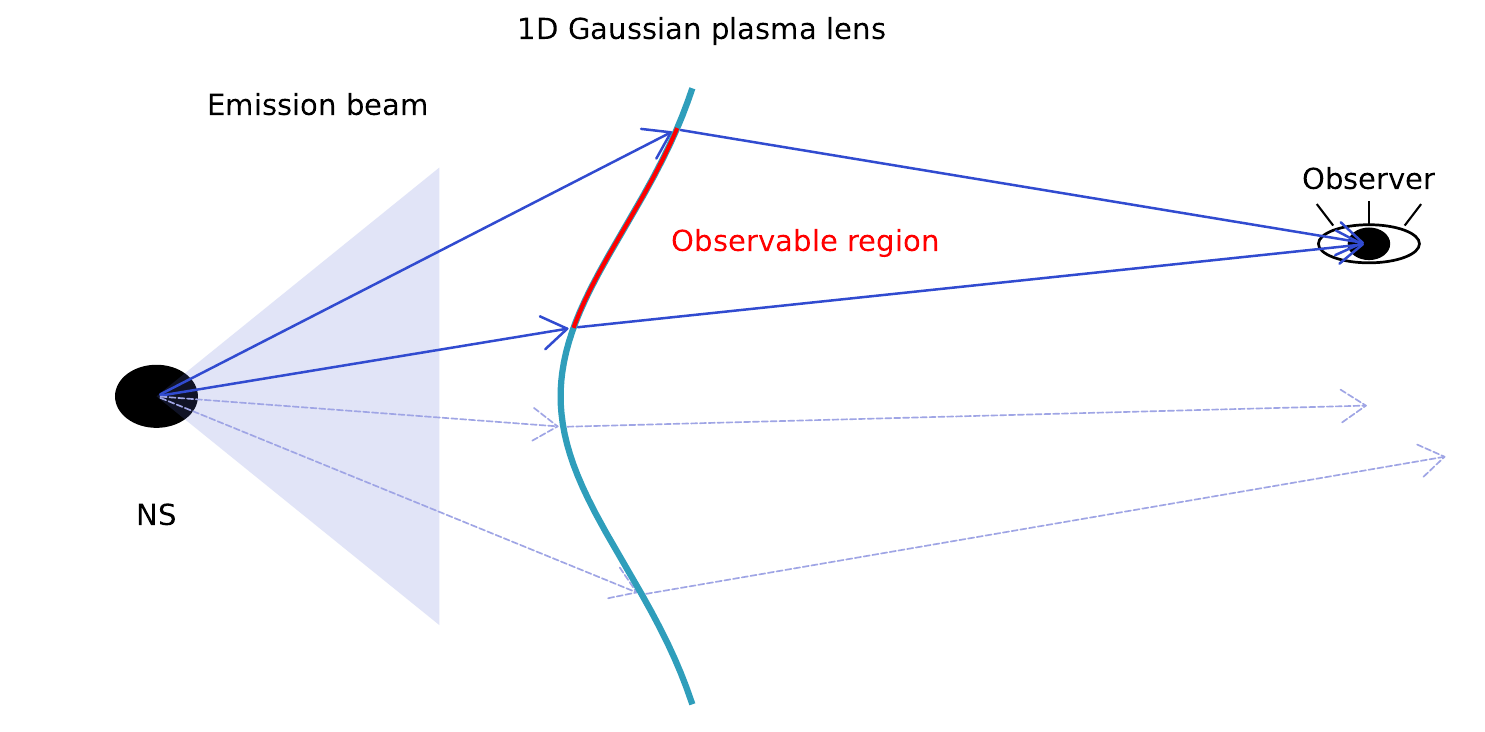}
\caption{Schematic illustration of a near-source 1D Gaussian plasma lens. A sweeping emission beam (caused by slow rotation or precession) intercepts a plasma lens with a transverse Gaussian electron-density profile. Only during part of the sweep phases, the rays pass through the caustic segment of the lens and are refracted toward the observer; at other phases, the rays miss the line of sight. In this configuration, the combination of the geometric sweep and the limited observable caustic region of plasma lens naturally gives rise to a frequency-dependent active window.
}
\label{Fig5}
\end{figure}

The schematic illustration of this scenario is presented in Figure \ref{Fig5}. The neutron star undergoes slow rotation or precession, during which its emission beam periodically sweeps across the plasma lens. Radio emission is detected by the observer only within a specific range of rotational phases, resulting in the observed periodic activity. If the observable emission beam traverses less than half of the lens plane and does not cross the lens center within the observable phase interval, a frequency-dependent active window can naturally emerge from the plasma lensing effect. 

\begin{figure} 
\centering
\includegraphics[width=100 mm]{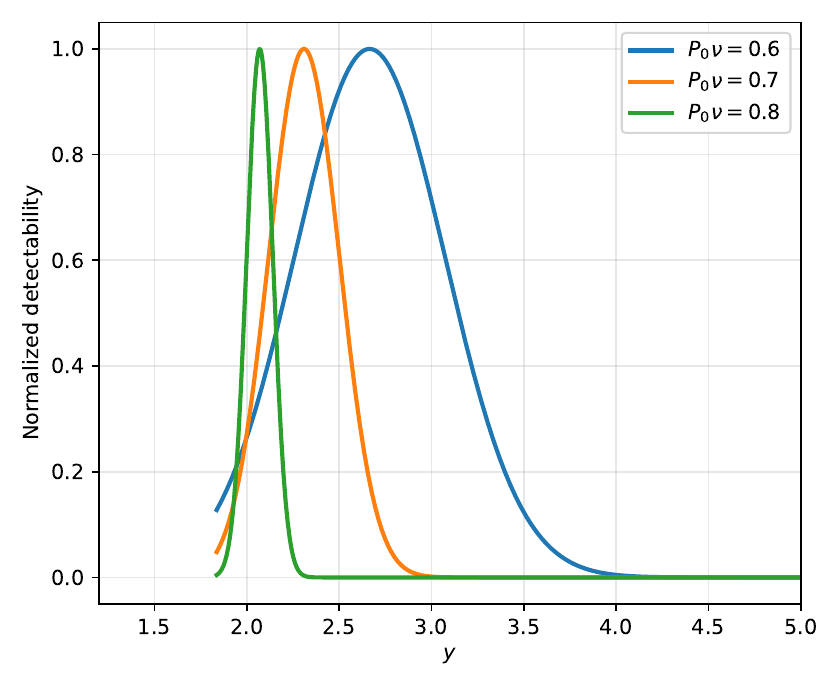}
\caption{Normalized detectability proxy as a function of transverse coordinate $y$ at fixed scaled frequency $s\equiv P_{0}\nu$.
For each scaled frequency $s=\{0.6,0.7,0.8\}$, we calculate the caustic-allowed intervals in $y$ from the intersections with the lower and upper caustic envelopes. Each interval [$y_{\rm low},y_{\rm high}$] is replaced by a Gaussian centered at $y_c=(y_{\rm low}+y_{\rm high})/2$ with FWHM set to $y_{\rm high}-y_{\rm low}$. The resulting curves provide a normalized representation in which the normalized detectability peak at each scaled frequency $s$ traces the phase drift of the active window, whereas the Gaussian width reflects the window’s duration.
}
\label{Fig6}
\end{figure}

As shown in Figure \ref{Fig1}, the lens enhances the received radio intensity within the caustic region, making the magnified bursts more likely to be detected than the unlensed ones. For a given scaled frequency $s \equiv P_0 \nu$, we determine the caustic-allowed phase interval in $y$ from the crossings $y_{\text {low}}(s)$ and $y_{\text {high}}(s)$. To visualize how an active window broadens or shifts with $s$, we simply replace each allowed segment by a Gaussian of width equal to the segment span ($\sigma=y_{\text {high}}-y_{\text {low}}$). As shown in Figure \ref{Fig6}, the Gaussian width encodes the caustic-window width and serves only as a smooth, readable proxy for detectability across phase.

To accurately reproduce the specific frequency–phase mapping seen in FRB 20180916B, a purely chromatic but constant scaled parameter $s \equiv P_{0}\nu$ is generally insufficient: a single $P_0$ cannot keep the caustic band both narrow at high $\nu$ and broad at low $\nu$, as required by the data. In practice, however, the observed frequency–phase structure is shaped not only by plasma lensing, but also by instrumental response, additional propagation effects, intrinsic spectral–temporal variability of the emission and the triggering and activity mechanisms of the FRB engine. As a consequence, a fully quantitative fit that reproduces all details of the dynamic spectrum is highly non-trivial, since multiple effects act simultaneously on the observed signal. In this work, we therefore limit our discussion to a qualitative level and, as an illustrative example, focus on one specific form of additional chromatic modulation. In particular, scattering-induced broadening of the lens can introduce extra chromatic dependence of $s(y,\nu)$, with the effective transverse scale evolving as:
\begin{equation}
a_{\mathrm{eff}}^2(\nu)=a_0^2+\sigma_0^2 \nu^{-2 m_\theta} 
\end{equation}
where $m_\theta \simeq2-2.2$ for a Kolmogorov cascade but may deviate from this value in realistic environments. This leads to a chromatic lens strength $\alpha(\nu) \propto \nu^{-2} / a_{\mathrm{eff}}^2(\nu)$. Such a mechanism introduces an additional frequency-dependent modulation to the phase–$y$ mapping (phase=$f(y)$), allowing the model to naturally produce both the early, narrow high-frequency active windows and the late, broader low-frequency windows seen in FRB 20180916B. 

\section{Constraint on the size of the emission region} \label{sec:emission size}
\begin{figure} 
\centering
\includegraphics[width=120 mm]{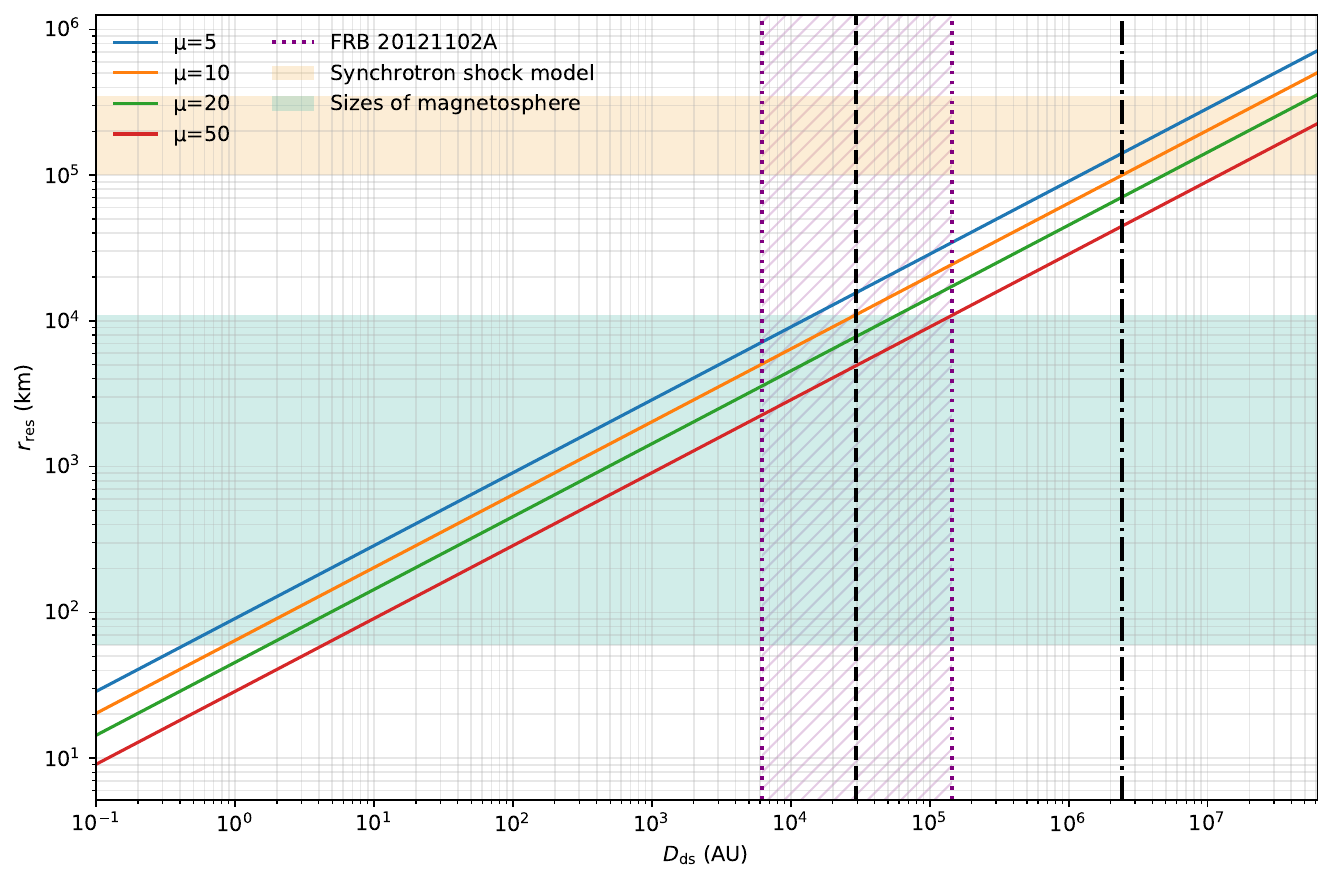}
\caption{1D plasma-lens resolution $r_{\mathrm{res}}$ as a function of the source–lens separation $D_{ds}$ at $\nu$ = 1.25 GHz.
Curves in different colors denote the values of $r_{\mathrm{res}}$ corresponding to flux magnifications of $\mu = \{5, 10, 20, 50\}$. The amber-shaded region marks the representative source size range predicted by the synchrotron shock model \citep{Margalit2020a}, while the teal-shaded region indicates the possible extent of the magnetosphere, estimated from the light-cylinder radius $R_{\mathrm{LC}}=c / 2 \pi f$. The upper and lower bounds are constrained by the slowest known pulsar \citep{Tan2018} and the fastest-spinning pulsar \citep{Hessels2006}, respectively. The vertical dashed line represents the upper limit of $D_{ds}$ required to distinguish between the inner- and outer-magnetospheric emission models of $\mu=10$. The vertical dashdot line represents the upper limit of $D_{ds}$ required to map the emission region within the magnetosphere under the same flux magnification. The purple hatched region denotes the possible range of the lens–source separation $0.03~\rm{pc}< D_{ds} < 0.7~\rm{pc}$ for FRB 20121102A, inferred from the RM/DM evolution and the properties of the associated PRS. 
}
\label{Fig7}
\end{figure}

FRB emission models are commonly grouped into magnetospheric scenarios (e.g., \citealt{Katz2014, Kumar2017, Katz2018, Lu2018, Yang2018, Cooper2021, Zhang2022_ICS, Qu2024}), in which coherent radiation is produced within or near the light cylinder, and outer-magnetospheric or shock-driven scenarios (e.g., \citealt{Lyubarsky2014, Beloborodov2017, Metzger2019, Plotnikov2019, Beloborodov2019, Margalit2020a, Wu2020}), where the emission arises at much larger radii. Distinguishing between these classes requires robust upper limits on the transverse size of the emission region. Several observational signatures already point toward magnetosphere-scale emission sizes (e.g. microsecond substructure \citep{Nimmo2022, Snelders2023}, polarization-angle evolution \citep{Luo2020, Niu2024, Liu2025, Mckinven2025}, and scintillation-based constraints \citep{Kumar2024, Nimmo2025}). To obtain complementary, geometry-anchored bounds with weaker microphysics assumptions, we exploit plasma-lensing resolution near the source to place direct limits on the emitting region. In the frame work of plasma lensing, a nearby lensing screen offers the strongest localization, with the constraint tightening as the lens approaches the source based on the wave optics \citep{cordes2017, Main2018}.
For a strong lensing event, the intrinsic source size must be smaller than the resolution of the 1D plasma lens \citep{Main2018}
\begin{equation}
r_{\mathrm{res}} \approx  1.9\sqrt{\frac{{\lambda a}}{\pi \mu }}.
\end{equation}
Some active repeaters may reside in binaries with orbital periods of several tens to a few hundred days \citep{Wang2022, LiYe2025}, implying semi-major axes of order AU. In that case, a circum-stellar lens near the companion naturally yields AU-scale $D_{ds}$ and correspondingly fine resolution. Alternatively, if the lens sits in a supernova remnant at $D_{ds}\sim$ pc or in a nebula at $D_{ds}\sim$ sub-pc from the magnetar, the achievable resolution is coarser. As a fiducial example, for $\mu=10$ and $\nu=1.25,\mathrm{GHz}$, one finds $r_{\mathrm{res}}\sim 60~ \mathrm{km}$ for $D_{ds}\sim1\ \mathrm{AU}$ and $r_{\mathrm{res}}\sim 3\times10^{4}\ \mathrm{km}$ for $D_{ds}\sim1\ \mathrm{pc}$. In Figure \ref{Fig7}, the vertical dashed line marks $D_{ds}\lesssim 11\ \mathrm{pc}$, within which plasma-lensing constraints can, in principle, distinguish inner- versus outer-magnetospheric emission scenarios. The vertical dash-dotted line indicates $D_{ds}\lesssim 0.1\ \mathrm{pc}$, within which mapping of emission within the magnetosphere becomes feasible. 

A possible plasma-lensing event has been reported in FRB 20121102A \citep{Platts2021}. Based on the maximum fluence contrast between adjacent bursts exhibiting complex time–frequency structure, a characteristic flux magnification of $\mu \sim 10$ at $\nu \sim 1.25~\mathrm{GHz}$ is reasonable \citep{Caleb2020}. For this source, the lensing structure may reside within the magnetar wind nebula (MWN), for example in the form of small-scale filaments. If the compact persistent radio source (PRS) associated with FRB 20121102A originates from synchrotron emission of the MWN, its inferred radius is $R_{\rm n} \simeq 0.4~\mathrm{pc}$ \citep{Zhao2021b}. Independent modeling of the time-varying DM and RM further suggests a source age of $\sim 9$–$10$ yr in the MWN scenario \citep{Zhao2021}. Adopting a characteristic ejecta velocity $v_{\rm n} \lesssim 10^{4}~\mathrm{km,s^{-1}}$ \citep{Margalit2018} yields a nebular radius of $R_{\rm n} \sim 0.1~\mathrm{pc}$, implying $D_{ds} \simeq R_{\rm n}$ within this range. VLBI measurements show that the physical extent of the co-located PRS is $\lesssim 0.7~\mathrm{pc}$ \citep{Marcote2017}, while scintillation constraints place a lower limit of $\gtrsim 0.03~\mathrm{pc}$ on this size \citep{Chen2023}. These measurements together imply $0.03~\mathrm{pc} \lesssim D_{ds} \lesssim 0.7~\mathrm{pc}$, consistent with the MWN-based expectations. Under these conditions, plasma-lensing constraints disfavor an outer-magnetospheric origin of the emission, as illustrated by the purple hatched region in Figure \ref{Fig7}. These constraints can even be used to localize the emission region on sub-magnetospheric scales.

\section{Discussion} \label{sec:dis}
Statistical studies have shown that downward-drifting patterns are observed much more frequently in active repeaters \citep{Zhou2022, ZhangLX2025}. In the framework of plasma lensing, the drift direction is set by the sign of the angular deviation $\delta\theta$, and in the absence of evidence for a preferred sign we assume that positive and negative values occur with equal probability. Within the outer caustic region, however, the frequencies at which the radiation is magnified is much more sensitive to changes in $y$ near the inner boundary than near the outer boundary as illustrated by the magnification map in panel (b) of Figure \ref{Fig1}. Since inward motion corresponds to upward drifting in our model, modest inward shifts in $y$ more readily move the magnified band out of the observing range or into the inner caustic, where the emission is strongly suppressed, so that the second burst often falls below the detection threshold. This observational selection effect naturally biases the detected sample toward downward-drifting events and, together with the truncation of the caustic at high frequencies, also makes drifting patterns intrinsically more difficult to detect at higher frequencies, consistent with broadband observations of FRB 20240114A \citep{Limaye2025}.
Moreover, drifting patterns arise when the emission passes through the caustic region of a plasma lens, where the flux magnification is strongest. Consequently, bursts showing such drifting features are expected to appear brighter than the unlensed ones. Nevertheless, the intrinsic intensity of FRB emission is highly variable from burst to burst and is further affected by the instrumental bandpass, which complicates direct comparison. Theoretically, both magnetospheric emission processes and plasma lensing in the surrounding medium could produce drifting patterns, making it difficult to distinguish between the two scenarios.

\cite{Qu2025} argued that plasma lensing is unlikely to explain the PA orthogonal jump occurred in FRB 20201124A due to the excessive cusp density required for three jump events in over 2000 bursts based on the point-like cusp assumption. However, FRB 20201124A may reside in a binary system \citep{Wang2022, LiR2025, XuJW2025}, in which case the relative velocity between the magnetar and the circum-stellar medium is the sum of the velocity of the magnetar proper motion and the stellar wind or the disk which can be as high as $v_{r}\sim10^8~\rm{cm~s^{-1}}$ \citep{Snow1981}. The corresponding linear density of the lenses can be reduced to less than the maximum value accquired in \cite{Qu2025}. Besides, the temporal separation of the first and second event ($\sim171$ day) is much larger than the potential orbital period (several tens of days) \citep{Wang2022, LiR2025, XuJW2025}. Therefore the number of the lens cusp could be less than three. In conclusion, plasma lensing can produce the observed orthogonal PA jump in FRB 20201124A under reasonable parameters. 

For most pulsars, the radio emission exhibits stable phase and intensity, and their spectra generally follow a power-law behavior with a negative spectral index. In contrast, active repeating FRBs display highly variable emission intensities and usually lack any clear periodicity, which makes the identification of plasma lensing events extremely challenging. This may explain why no conclusive plasma lensing event has been reported to date. Nevertheless, bursts with complex time–frequency structures and frequency-dependent dispersion measures across sub-bursts have been observed in many sources, including both repeaters and apparently non-repeating FRBs (e.g., \citealt{Gajjar2018, Marthi2020, Platts2021, Zhou2022, Kumar2023, Faber2024, ZhangLX2025}). Such complex features may represent the footprints of plasma lensing, and the distinct frequency dependence of arrival-time delays can be used to search for and verify the presence of plasma lenses \citep{Er2022}.

\section{Summary} \label{sec:con}
In this work, we systematically investigate the potential of a 1D plasma lens model to account for several observational properties of FRBs which is summarized as follows: 
\begin{enumerate}
    \item Small, monotonic variations in the incidence angle of the FRB wavefront across the lens cause the effective lens coordinate $y$ to evolve with time, shifting the caustic-amplified frequency band and naturally producing both downward and upward sub-burst drifts. This mechanism also explains why drifting features are more prominent at lower frequencies and downward drifts are more frequently observed, consistent with current observations.
    
    \item A near-source plasma lens can reproduce the orthogonal PA jumps observed in FRB 20201124A: when lensed images with different RMs and comparable intensities overlap, their polarization vectors become nearly orthogonal, generating an apparent $90^\circ$ transition accompanied by a reduction in $L/I$.
    
    \item The chromatic active window of FRB 20180916B arises naturally when the emission beam of a slowly rotating neutron star sweeps across an asymmetric plasma lens, so that only certain rotational phases satisfy the caustic-amplification condition, yielding earlier and narrower activity windows at higher radio frequencies and later, broader windows at lower frequencies.
    
    \item Plasma lensing events can provide constraints on the transverse emission size, with the achievable spatial resolution strongly dependent on the source-lens separation $D_{ds}$. Near-source lenses, such as those in the stellar wind/disk of companion, nebula or SNR, can probe emission structures on magnetospheric scales, whereas more distant lenses provide coarser limits. For FRB 20121102A, the inferred $D_{ds}\sim0.03-0.7$ pc disfavors an outer-magnetospheric emission origin.
\end{enumerate}

These results demonstrate that plasma lensing offers a plausible and unifying framework for several complex observed behaviors and may play a significant role in FRB observations.

\section*{Acknowledgments}
This work was supported by the National Natural Science Foundation of China (grant Nos. 12494575, 12273009 and 12403054), the National SKA Program of China (grant No. 2022SKA0130100), the Natural Science Foundation of Xinjiang Uygur Autonomous Region (No. 2023D01E20), the Natural Science Foundation of Sichuan Province (No. 2025ZNSFSC0878).

\bibliographystyle{aasjournal}
\bibliography{ref1}

\clearpage

\end{document}